\documentclass[journal]{IEEEtran}
\usepackage{cite}
\usepackage{amsmath,amsfonts}
\usepackage{algorithmic}
\usepackage{algorithm}
\usepackage{array}
\usepackage{hyperref}
\usepackage[caption=false,font=normalsize,labelfont=sf,textfont=sf]{subfig}
\usepackage{textcomp}
\usepackage{stfloats}
\usepackage{url}
\usepackage{verbatim}
\usepackage{graphicx}
\usepackage{soul,color}
\usepackage{siunitx}
\usepackage{tablefootnote}
\usepackage{multirow}
\usepackage{booktabs}
\usepackage{tabularray}
\usepackage{bbm}
\usepackage{tikz}

\newcommand*\circled[1]{\tikz[baseline=(char.base)]{
            \node[shape=circle,draw,inner sep=0.5pt] (char) {#1};}}

\begin{document}

\title{Leveraging Visual Supervision for Array-based Active Speaker Detection and Localization}

\author{Davide Berghi,~\IEEEmembership{Student Member,~IEEE},
and Philip J.B. Jackson,~\IEEEmembership{Member,~IEEE}

\thanks{The authors are with the Centre for Vision, Speech and Signal Processing (CVSSP) of the University of Surrey, Guildford, Surrey GU2 7XH, U.K. (e-mail: \{davide.berghi, p.jackson\}@surrey.ac.uk).

}%
}

\markboth{IEEE/ACM Transactions on Audio, Speech, and Language Processing}%
{Berghi \MakeLowercase{\textit{et al.}}: Multichannel Audio Active Speaker Detection and Localization using Visual Supervision}


\maketitle

\begin{abstract}

Conventional audio-visual approaches for active speaker detection (ASD) typically rely on visually pre-extracted face tracks 
and the corresponding single-channel audio to find the speaker in a video.
Therefore, they tend to fail every time the face of the speaker is not visible.
We demonstrate that a simple audio convolutional recurrent neural network (CRNN) trained with spatial input features extracted from multichannel audio can perform simultaneous horizontal active speaker detection and localization (ASDL), independently of the visual modality. 
To address the time and cost of generating ground truth labels to train such a system, we propose a new self-supervised training pipeline that embraces a ``student-teacher'' learning approach. A conventional pre-trained active speaker detector is adopted as a ``teacher'' network to provide the position of the speakers as pseudo-labels. The multichannel audio ``student'' network is trained to generate the same results. At inference, the student network can generalize and locate also the occluded speakers that the teacher network is not able to detect visually, yielding considerable improvements in recall rate.
Experiments on the TragicTalkers dataset show that an audio network trained with the proposed self-supervised learning approach can exceed the performance of the typical audio-visual methods and produce results competitive with the costly conventional supervised training. 
We demonstrate that improvements can be achieved when minimal manual supervision is introduced in the learning pipeline. 
Further gains may be sought with larger training sets and integrating vision with the multichannel audio system.


\end{abstract}


\section{Introduction}
\IEEEPARstart{S}{ound} and vision present different yet complementary sensory information. On the one hand, visual sensors such as cameras, typically operate in the frontal field of vision (FoV) and cannot sense occluded objects, whereas the audio's field of audition (FoA) is omni-directional and does not require direct line-of-sight with the target. 
On the other hand, vision-based algorithms usually guarantee higher spatial resolution compared to audio systems.
A good audio-visual (AV) system must be designed to take full advantage of these complementary features but also to rely on a single modality when its counterpart is corrupted or fails.
This work investigates whether these systems can benefit from joint AV training to gain performance when the visual input fails. 

In particular, we consider the active speaker detection and localization (ASDL) task, i.e., detecting and locating the active speaker within the visual reference frame.
This task is vital as it relates to many other AI tasks, such as speaker diarization \cite{Gebru:2018:SpeakerDiarization}, human-robot interaction (HRI) \cite{He:2018:deepNetsHRI}, human activity recognition \cite{Gao2020ListenTL}, scene understanding \cite{Tian2018AudioVisualEL}, augmented reality (AR) \cite{jiang2022egocentric} and immersive media production \cite{Schweiger:2022:tool6dof}. 
Therefore, it has a huge breadth of potential applications in, for example, entertainment, communications, human-robot collaborative manufacturing, assisted living in health and social care, and other applications besides.

\begin{figure}[tb]
\centerline{\includegraphics[width=\columnwidth]{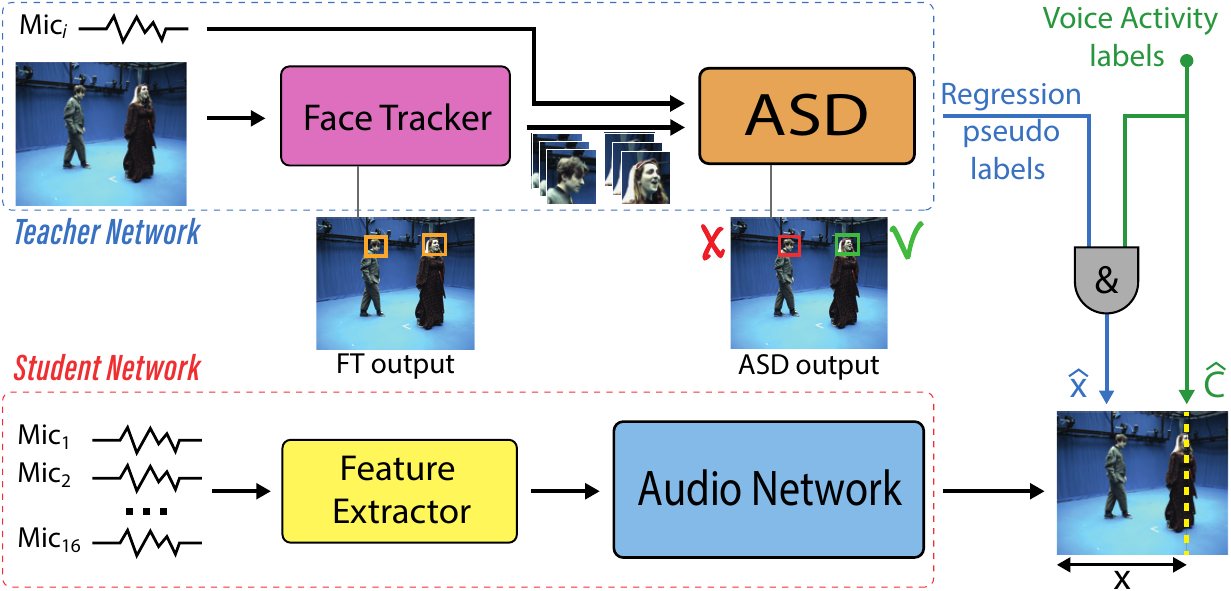}}
\caption{Block diagram of the proposed learning pipeline. The `teacher' network, composed of an audio-visual active speaker detector (ASD) preceded by a face tracker, produces pseudo-labels with the speaker's position ($\hat{x}$) at each time instance. The `student' audio network is trained to regress the speaker's horizontal position ($x$) from the directional audio features extracted from the multichannel input. Voice activity annotations are used to supervise the training for the detection subtask and to mask the teacher's pseudo-labels when the frame is silent.}
\label{block_diagram}
\end{figure}

ASDL can be tackled in two separate stages. The first stage corresponds to the localization task which is accomplished in advance by a visual face detector that builds a set of hypothetical candidate speakers. 
The second stage classifies the detected faces as active/inactive. This classification stage is called active speaker detection (ASD) and is typically performed as an AV problem with single-channel audio \cite{Roth:2020:AVA,Chung2019NaverAA,Alcazar_2020_CVPR,zhang2021unicon}.
In practice, the face of the speaker might not be visible from the viewpoint of the camera and therefore not be included in the set of candidates, causing the failure of the overall ASDL system.
We solve this problem 
holistically with an audio-only system that tackles the task using multichannel audio. 
Our system performs the detection and horizontal localization of the speaker in the visual frames simultaneously thanks to directional cues encoded in the multichannel audio signals. 
However, annotating the ground truth speaker positions to train the system, as is required in traditional supervised machine learning, is expensive and time-consuming. Therefore, we propose a self-supervised student-teacher training approach that leverages automatic \textit{pseudo-labels} generated by a conventional AV active speaker detector to supervise the training of our multichannel audio-based network. 
To the best of our knowledge, this work is the first attempt to develop an audio ASDL solution with automatic audio-visual supervision.
Experiments on the TragicTalkers dataset \cite{Berghi:2022:TragicTalkers} show that the visual modality can be employed as a helpful tool to automatically provide supervision for the proposed task, purely from multichannel audio signals. We demonstrate that, although only the visible speakers are employed to generate the supervisory signal used to train the audio network, the model is able to generalize at inference and detect also the occluded ones.
Furthermore, we show how our self-supervised learning pipeline can be augmented with little manual supervision. We demonstrate that it is possible to achieve results on par with the fully supervised case with considerably less labor-intense forms of supervision.
Our study presents the following three key contributions: 
\begin{itemize}
    \item For the first time, we leverage multichannel audio to simultaneously detect and horizontally locate the active speaker in the video frame domain, instead of state-of-the-art audio-visual solutions that rely on face detection. Additionally, we show via an ablation study that spatial feature extraction and temporal modeling are essential to properly tackle this task with multichannel audio;
    \item We propose a cross-modal self-supervised student-teacher learning approach in which a multichannel audio network is trained with localization pseudo-labels generated by a pre-trained audio-visual teacher network. We test two different teacher networks as well as hybrid versions of manual-automatic labels to find an optimal compromise for high performance and minimal human supervision;
    \item Experimental results on the TragicTalkers dataset demonstrate that the student network, trained on the visible active speakers detected by the teacher network, can detect occluded faces and generalize the localization, achieving higher recall and overall performance than its teacher.
\end{itemize}

An early version of our work was presented in \cite{Berghi:2021:mmsp}. 
Here we present an overhauled audio network with extended temporal modeling.
We adopt state-of-the-art audio input features in contrast to the beamformer-based features in \cite{Berghi:2021:mmsp}. 
The dataset size has been increased by $\sim$50\% with TragicTalkers and voice activity labels added. We now compare multiple teacher networks and a spectrum of supervision conditions.
A substantial performance improvement on this larger test set is achieved.       

To encourage reproducibility and motivate future research, our code is made available to the community\footnote{\href{https://github.com/dberghi/Leveraging-Visual-Supervision-for-Array-based-ASDL}{https://github.com/dberghi/Leveraging-Visual-Supervision-for-Array-based-ASDL}}. 
The remainder of the paper is organized as follows:
\ref{sec:background} presents the background; \ref{sec:method} describes in detail the proposed student-teacher learning method; \ref{sec:settings} the experimental settings and the experiments conducted; \ref{sec:results} presents and discusses the results and future work; \ref{sec:concl} concludes the paper.

\section{Background} \label{sec:background}

The task and the learning method proposed in this paper can be linked to three main research areas that are described below: active speaker detection (ASD), sound event localization and detection (SELD), and self-supervised audio-visual learning.

\subsection{Active Speaker Detection} \label{ASD_sebsec}

As shown in the upper half of Fig.\,\ref{block_diagram}, in the teacher network, ASDL is commonly treated as a two-step process: first a video-only face detector is applied to find and locate faces in the image sequence, second the detected face tracks are classified as active/inactive. This second subtask is termed active speaker detection (ASD), which typically forms an AV solution employing a single (mono) audio signal.
Therefore, ASD consists in identifying the presence of active speakers in a video among a set of candidates. 

Early works in this area attempted to find the correlation between voice activity and lip or upper body motion \cite{Cutler:2000:lookWho,Haider:2012:towardsspeaker,Chakravarty2015WhosSA}. 
In some cases, researchers proposed self-supervised solutions to perform ASD, e.g., by training a visual network under the supervision of its audio counterpart \cite{Chakravarty2015WhosSA}, or by audio-visual co-training \cite{Chakravarty2016ActiveSD,Hoover2019ICASSP}.
What is probably the first, large, annotated dataset for ASD was released for the ActivityNet Challenge (Task B) at CVPR 2019: the AVA-ActiveSpeaker dataset \cite{Roth:2020:AVA}. 
It provides 38.5 hours of audio-visual face tracks (sequences of consecutive face crops) labeled for speech activity.
Chung \textit{et al.} tackled the challenge with an AV model pre-trained on audio-to-video synchronization, performing 3D convolutions \cite{Chung2019NaverAA}. Alcázar \textit{et al.} proposed Active Speakers in Context (ASC) \cite{Alcazar_2020_CVPR}. Instead of compute-intensive 3D convolutions or large-scale AV pre-training, ASC uses context. 
Zhang \textit{et al.} have also tackled the ASD task by using contextual information and proposed the Unified Context Network (UniCon) \cite{zhang2021unicon}. 
Leveraging short- and long-term features and AV cross-attention, Tao \textit{et al.} introduced TalkNet \cite{tao2021TalkNet}. Additionally, motivated by the call for an ASD system that works properly outside the AVA-ActiveSpeaker dataset domain, they formed a second ASD dataset based on LRS3 \cite{Afouras:2018:LRS3} and VoxCeleb2 \cite{Chung:2018:voxceleb2} called TalkSet. 
Recently, Alcázar \textit{et al.} \cite{Alcazar2022EndtoEndAS} proposed an end-to-end ASD that unifies AV feature extraction and spatio-temporal context aggregation.
Although highly effective, these solutions simply perform the audio-visual classification of the provided pre-extracted face tracks. In practice, the speaker can be occluded or facing away from the camera, causing face detection to fail and degrade overall system performance. In other words, the active speaker is detected only when visible.

The approach proposed in this study compensates for this problem by extending the audio input front-end and employing a microphone array for localization: at inference, the model relates voice activity to the position of the speaker in the visual frame through the audio modality only. Therefore, the proposed solution simultaneously performs both detection and localization. 
Additional AV studies using multichannel audio signals in the field of ASDL learning include the work from Qian \textit{et al.} \cite{Qian:2021:AVFusion}, where visual feature vectors encoding face bounding box coordinates were used alongside audio features to improve spatial accuracy, and Jiang \textit{et al.} \cite{jiang2022egocentric}, where AV cues are used to locate the speakers within an egocentric environment for augmented reality (AR) applications. Other AV solutions employ multi-view and microphone arrays for 3D speaker detection and tracking and are typically based on classical (other than DNN-based) approaches \cite{izhar:2020:AVtracker,Qian:2019:AVSensDevic,Liu:2018:MultipleSpeakTracking,gebru2015tracking}. However, they often rely on acoustic and optical calibration data and require a computationally-expensive global search \cite{Do:2007:RT_SRP-PHAT}. Additionally, learning-based solutions with appropriate training have been shown to better generalize in highly reverberant environments and with low signal-to-noise ratio (SNR) conditions \cite{Xiao:2015:learning-based}. 

In this work, an audio network that takes in input acoustic features extracted from multichannel audio sequences has been trained to perform ASDL. In order to build a deep understanding of the task and the proposed components, the problem analyzed in this report is limited to the case of a single active speaker and horizontal localization.

\subsection{Sound Event Localization and Detection}

The audio ASDL task can be thought of as a specialization of sound event localization and detection (SELD)  \cite{Adavanne:2019:SELDnet} considering only speech and silence, defined purely in the audio modality. SELD intersects  two main subtasks, sound event detection (SED) and direction of arrival (DoA) estimation: SELD  simultaneously recognizes the target sound class, with its onset and offset times, while estimating its DoA. SELD methods provide relevant inspiration for the proposed method. 

SELD was introducted into the 2019 DCASE challenge as Task 3 to advance state-of-the-art methods \cite{Adavanne2019_DCASE}. 
The dataset used in the challenge provides sound scenes in two spatial formats: 4-channel First-Order Ambisonics (FOA) and 4-channel microphone array (MIC). 
In 2018, Adavanne \textit{et al.} pioneered the task by proposing SELDnet \cite{Adavanne:2019:SELDnet}, a convolutional recurrent neural network (CRNN) with magnitude and phase spectrogram inputs from the array's channels. 
A two-stage learning strategy was adopted by Cao \textit{et al.} \cite{Cao:2019:polyphonic} to decompose the problem into its SED and DOAE subtasks. 
They extracted generalized cross-correlation with phase transform (GCC-PHAT) spatial features alongside log-mel spectrograms as inputs for their two-stage network. Nguyen \textit{et al.} \cite{Nguyen:2020:SMN} also performed SED and DOAE separately, employing a Sequence Matching Network (SMN) to learn the correct match between the two subtasks' outputs. Activity-Coupled Cartesian Direction of Arrival (ACCDOA) vector-based loss unifies the regression and detection loss terms \cite{Shimada:2021:ACCDOA}. The loss for SELD typically sums the weighted SED and DOAE losses, whereas, ACCDOA assigns sound event activity to the DOA vector's magnitude. 
To solve the problem of simultaneous same-class events, Cao \textit{et al.} proposed the Event Independent Network (EIN) to generate independent track-wise predictions \cite{cao:2020:EIN}. Later, in EINv2 \cite{cao:2021:EINv2}, they replaced the bidirectional gated recurrent units (biGRUs) with multi-head self-attention (MHSA) and introduced soft parameter sharing between the SED and DOAE branches. Recently, Nguyen \textit{et al.} \cite{Nguyen:2021:SALSA} proposed a new type of spatial input feature for SELD: the Spatial Cue-Augmented Log-Spectrogram (SALSA) features. SALSA consists of a normalized version of the principal eigenvector of the array's spatial covariance matrix computed across microphone channels at each time-frequency bin of the spectrograms. 
A faster lightweight variation of SALSA called SALSA-Lite \cite{Nguyen:2021:SALSALiteAF} uses a frequency-normalized version of the inter-channel phase difference (NIPD).

\subsection{Self-Supervised Audio-Visual Learning}

Self-supervised AV learning is a research area that is rapidly gaining interest across the audio and visual communities. It provides not only multi-modal solutions to tackle traditional problems, e.g., exploiting 
visual information for speech enhancement and separation or solving the `cocktail party' problem \cite{Afouras18,Ephrat:2018:Looking2Listen}, but
it has also given rise to new tasks: e.g., Morgado \textit{et al.} \cite{morgadoNIPS18}, Gao \textit{et al.} \cite{Gao:2019:visualsound}, and Yang \textit{et al.} \cite{Yang:2020:CVPR} proposed self-supervised approaches to generate spatial audio from videos with monaural sound.
Other researchers performed localization or separation of audio sources that were seen in the video \cite{Arandjelovic2017ObjectsTS,gao2018objectSounds,Gao:2019:coseparating,zhao:2018:ECCV}.
These approaches rely on the natural co-occurrence of audio and visual events to learn a common audio-visual representation, e.g., to associate the sound of a guitar with its visual appearance.
However, the localization task is performed visually. These approaches require both modalities present, which might fail with poor lighting or visual occlusion. 
In this paper, the visual modality is employed only during training, drawing on the student-teacher paradigm \cite{Hinton2015DistillingTheKnowledge,Aytar:2016:soundNet,Owens2016AmbientSP,Albanie:2018:emotion}. 
In AV learning, this paradigm typically exploits the natural synchronization of audio and visual signals to bridge modalities, enabling one to supervise its counterpart.
Owens \textit{et al.} \cite{Owens2016AmbientSP} used audio to supervise and improve visual learning, while Aytar \textit{et al.} \cite{Aytar:2016:soundNet} used vision to supervise audio learning. Elsewhere, audio and vision supervise each other \cite{Arandjelovic:2017:look}.
Closest to the proposed work, \cite{Gan2019SelfSupervisedMV,valverde:2021:mmdistillnet,vasudevan:2020:semanticobject} adopted a student-teacher approach to detect vehicles in the visual domain using multichannel audio input. The audio student models are trained with 2D bounding boxes generated by a pre-trained visual teacher network: 
Gan \textit{et al.} \cite{Gan2019SelfSupervisedMV} and Rivera Valverde \textit{et al.} \cite{valverde:2021:mmdistillnet} used respectively a stereo microphone and a microphone array to estimate vehicle locations, while Vasudevan \textit{et al.} \cite{vasudevan:2020:semanticobject} used binaural sound for semantic segmentation of 360° street views. 
Here, we use a 16-element microphone array and aim to detect human speakers, not cars. 
While the engine of a moving vehicle emits a continuous source of sound, speech is intermittent and the detected face can be actively speaking or silent. Thus, speech activity is also considered in our approach below.


\section{Student-Teacher Learning Method} \label{sec:method}

The proposed learning approach is both self-supervised and semi-supervised \cite{LeCun:2019:tweet,Zhu:2009:semiSupervised}. It consists of a single-channel audio-visual teacher network and an audio-only student network as in Fig.~\ref{block_diagram}. 
ASDL comprises classification and regression. Each frame is classified as active or silent and, when active, the position of the speaker is regressed.
The student network is trained to perform these two subtasks concurrently. 
In contrast, the teacher network first visually detects and tracks faces in the video, then employs an AV model pre-trained on ASD to classify the face tracks with the help of the corresponding single-channel audio. 
The horizontal positions of the center of the bounding boxes of the active faces are then used as location \textit{pseudo-labels} to train the student network, which employs multichannel audio from a microphone array.
In terms of detection, the pseudo-labels present an essential problem: when no speaker is detected, it is not possible to ensure that the cause is the effective absence of speech or a missed detection caused by visual occlusion. Thus, the teacher's negative detections are not reliable, 
so we label the audio data for voice activity (VA) to supervise detection. In the self-supervised pipeline, \textit{VA labels} are provided by an automatic voice activity detector (VAD).  

Even though only the positions of the visible speakers are used to supervise the regression subtask, at inference the student network can take full advantage of the spatial soundfield captured by the microphone array to generalize to occluded speakers.
To implement this method, a dataset of AV speech data is required where the soundfield is captured with a microphone array. 
We now describe details of the dataset, teacher and student networks, and loss function used to combine location and VA labels.

\begin{figure}[tb]
\centerline{\includegraphics[width=\columnwidth]{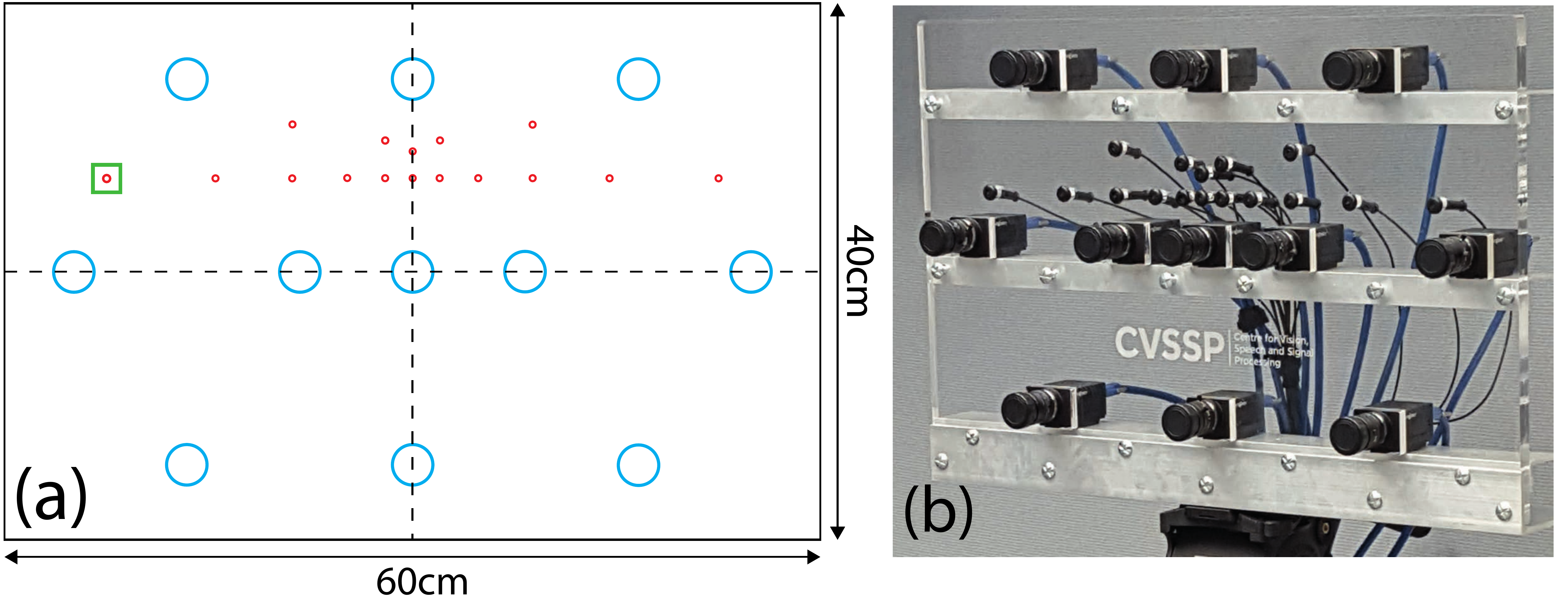}}
\caption{(a) Schematic of camera (blue circles) and microphone (red dots) positions on the AVA Rig. The green square highlights the reference microphone. (b) Photo of an AVA Rig.}
\label{AVArig}
\end{figure}

\subsection{Multichannel AV Dataset}

The AV dataset used to train and test the proposed method is TragicTalkers \cite{Berghi:2022:TragicTalkers}. It offers sequences captured with the aid of twin Audio-Visual Array (AVA) rigs. Each AVA rig is a light-field and sound-field sensing platform consisting of a 16-element microphone array and 11 cameras fixed on a flat perspex baffle as in Fig.\,\ref{AVArig}. 
As the proposed learning pipeline requires a single video feed with the microphone array, the availability of multiple views allows us to extend the audio network training by selecting each camera view via a one-hot vector.
Hence, the network learns the array's mapping to the selected camera view.
Sequences include one or two actors at a range of approximately 3--4\,m performing monologues, conversations, and interactive scenes in which they move and occlude each other. The dataset was captured in an acoustically treated laboratory with an average reverberation time of 0.3s in the mid 0.5-2\,kHz frequency range. It does not contain sequences in which the speakers talk simultaneously, off-screen talkers, or external sources of sound other than speech. The background noise floor is minimal (SNR\,$\geq$\,30\,dB). 
Studying at most one active talker allows rigorous assessment of the proposed learning method in a studio environment as a realistic media production setting. Nevertheless, to assess the robustness of the audio network, we also conducted experiments with additive pink noise.
The actors cover the entire image FOV (2448p, [$\pm$27.5°]) distributed with 5th percentile at $-$15.7° (528p), and 95th percentile at $+$14.1° (1851p). They are separated by an angular distance that varies from 0°, when one occludes the other, to a maximum of 34.3°, with a median of 16.4°. We verified that the actors' angular separation has negligible influence on localization accuracy, since their speech does not overlap (Pearson correlation $<$\,0.1). 
TragicTalkers consists of 30 scenes captured with two AVA rigs. 
{We use the AV streams of each rig independently, i.e., 16-channel audio is used to predict the speaker's position in any one of the rig's 11 camera views. 
So the dataset's 30 scenes provide 60 rig sequences, each with 11 perspectives.} 
TragicTalkers offers ground truth (GT) labels for VA and 2D face bounding box. We use the GT labels to compare the proposed learning approach with the traditional supervised one.
The dataset is partitioned into a 50-sequence development set and a 10-sequence test set, which includes mouth position labels for evaluation.

\subsection{Teacher Networks} \label{chap:teacher}

The teacher network automatically creates pseudo-labels with the position of the speakers in the video frames. It consists of two main components: a face tracker and an ASD model. 
In this study, we evaluate two teacher networks.
For the first teacher, the face tracks (i.e., stacks of face crops) are generated with the SeetaFaceEngine2 face detector \cite{wu2016Seeta} applied to each TragicTalkers video frame.
SeetaFaceEngine2 was adopted as it proved to be 
effective and fast: it processes 2448$\times$2048p frames at 10$^+$\,fps on a 6-core Intel i7-9750H CPU. 
The per-frame detections are temporally tracked based on bounding-box intersection over union (IoU) across adjacent frames, and a Gaussian smoothing is applied. 
We train the publicly-available ASC model by Alcázar \textit{et al.} \cite{Alcazar_2020_CVPR} on the AVA-ActiveSpeaker dataset \cite{Roth:2020:AVA}. 
Then, it is used to classify the active speakers from the face tracks. 
When a speaker is detected, the horizontal position of the bounding-box center is used to supervise the audio student network in the regression subtask. We briefly call this teacher network \textsc{``Asc''}.
The second teacher network is based on the TalkNet model by Tao \textit{et al.} \cite{tao2021TalkNet}. They provide an end-to-end ASD demo to detect faces and classify them with a pre-trained TalkNet model. We call this teacher 
\textsc{``TalkNet''}. 
It employs S3FD \cite{Zhang2017S3FDS} as face detector and the classifier is trained on TalkSet, a dataset based on VoxCeleb2 \cite{Chung:2018:voxceleb2} and LRS3 \cite{Afouras:2018:LRS3}, where the faces are extracted with S3FD too.
While both classifiers work 
well on the AVA-ActiveSpeaker dataset (mAP results are 86.7 for ASC, 90.8 for TalkNet),
ASC does not generalize well to videos in the wild. 
A matching face tracker for training and testing would generalize better, giving the classifier a consistent facial area, but this is hard to re-implement. 
In contrast, aiming for general audio-visual ASD, Tao \textit{et al.} aligned the face tracker with TalkSet. 
Thus, \textsc{TalkNet} achieves 82\% average precision (AP) on the test set of TragicTalkers, while \textsc{Asc} only 59\%. 
To study the effects of teacher quality on the proposed student-teacher pipeline, both weak and strong teacher networks are assessed.

\begin{figure}[tb]
\centerline{\includegraphics[width=\columnwidth]{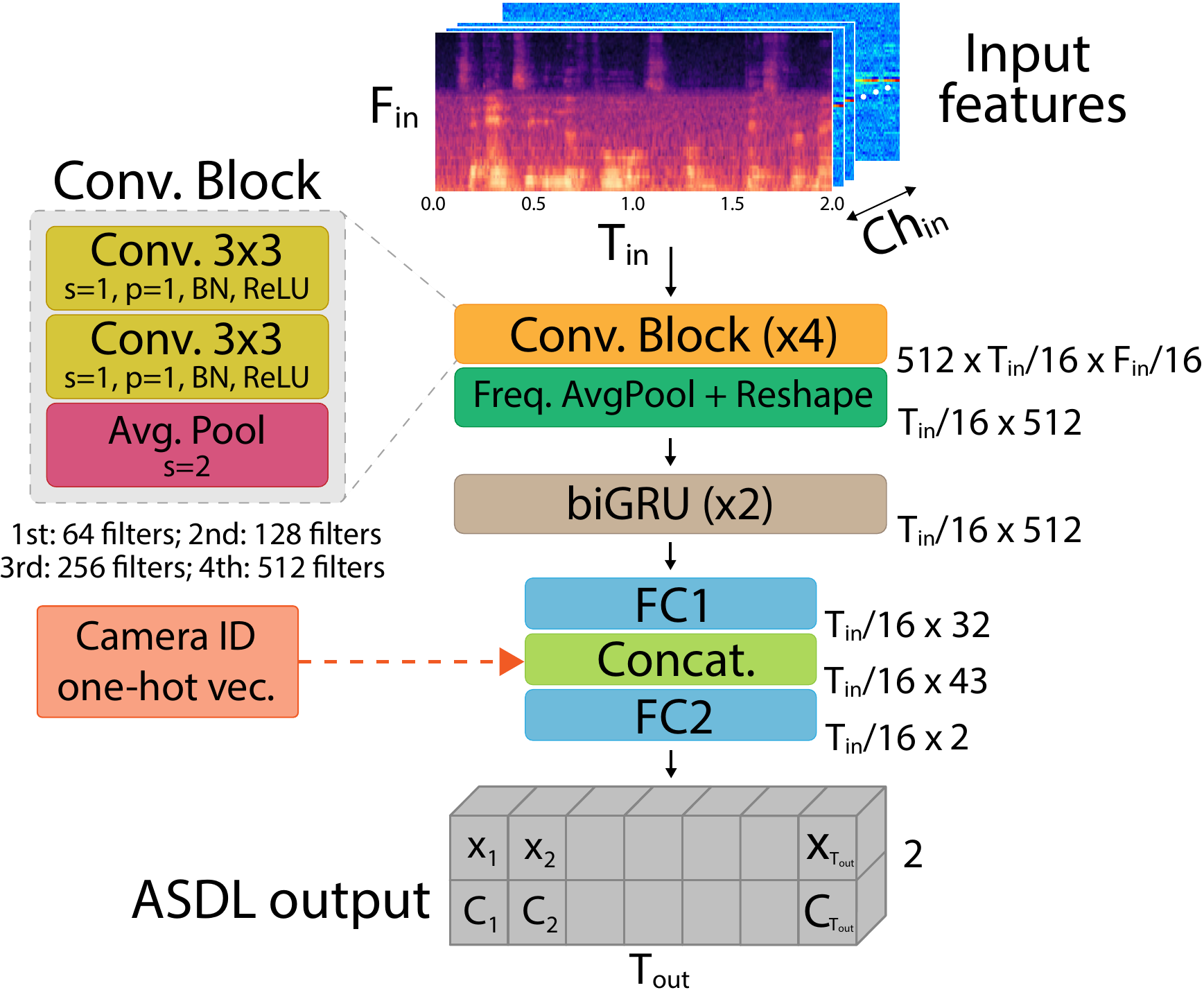}}
\caption{Schematic representation of the audio network architecture.}
\label{audio_network}
\end{figure}

\subsection{Student Network} \label{student_net_section}

The audio student network is trained under the supervision of the teacher network and the VA labels.
The network takes spatial features extracted from the multichannel audio. The input features have shape $Ch_{in}\times T_{in}\times F_{in}$, corresponding to the number of channels, time bins, and frequency bins, respectively. The number of input channels in the first convolutional layer is consistent with  $Ch_{in}$.
A CRNN-based architecture is adopted as depicted in Fig.\,\ref{audio_network}. We decided to opt for an existing backbone architecture. So, inspired by \cite{Cao:2019:polyphonic}, we used four convolutional blocks composed of two 3$\times$3 convolutional layers and an average pooling layer each. Both convolutional layers in the block are followed by batch normalization \cite{ioffe:2015:BN} and ReLU activation function. The average pooling layer is applied with stride\,$=$\,2 so that both time and frequency resolutions are halved after each convolutional block, while the number of channels is increased. At the end of the fourth block, the resulting tensor has shape $512\times T_{in}/16\times F_{in}/16$. Then, frequency average pooling is applied and the tensor is reshaped to $T_{in}/16\times 512$, as a series of 512-dimensional feature maps, one per video frame. After that, the features are passed through two bidirectional gated recurrent units (biGRUs) with 256 hidden units to learn two-way correlations over the sequences of frames. The output shape is preserved. Finally, the tensor is fed to two fully connected (FC) layers that reduce the feature map dimensionality to 2 output predictions per time frame: the regression value $x_i$ and the respective speech activity confidence $C_i$. 
The output coordinates and confidence are normalized in the range [0, 1] by a Sigmoid activation function. 
When the confidence $C_i$ is below a threshold, the $i$-th frame is considered silent and $x_i$ is neglected.   
Between the two FC layers, an 11-dimensional one-hot vector encoding the camera view is concatenated to each feature map.
$T_{in}$ is chosen so that the output rate of $T_{out}=T_{in}/16$ matches the frame rate.
The concatenation is done at this stage of the network where the feature map length approximates that of the one-hot vector.

\subsection{Loss Function}

The loss function has two terms: a regression loss and a confidence loss.
The supervisory signal used to compute the regression loss is generated from the positive predictions of the teacher network, i.e., the positions of the detected speakers. 
As mentioned previously, negative predictions do not guarantee absence of speech as they can occur when the face is occluded. 
The per-frame VA labels from the audio serve to disambiguate. Hence, a sum-squared error loss is computed at each output frame to train the student network:
\begin{equation}
\mathcal{L}=\sum_{i=1}^{T_{in}/16}\mathbbm{1}_i(x_i-\hat{x}_i)^2+(C_i-\hat{C}_i)^2
\label{loss_function}
\end{equation}
where $x_i$ and $\hat{x}_i$ are respectively the predicted and target positions of the speaker along the horizontal axis of the $i$-th video frame, normalized in the range [0, 1], while $C_i$ and $\hat{C}_i$ are the predicted and target confidences. The confidence loss is trivially achieved using the VA annotations: $\hat{C}_i$ is set to 1 when the frame is active and 0 when silent.
The masking term $\mathbbm{1}_i$ is 1 only when VA is positive \textit{and} the speaker is detected by the teacher network. It is set to 0 otherwise. So, when the frame is silent, or when the frame is active but the teacher's pseudo-label is unavailable, the network is only penalized by the confidence loss and not by the regression loss. The target position of the speaker, $\hat{x}_i$, is provided by the teacher network's location pseudo-labels.

\section{Experimental Methodology} \label{sec:settings}

\subsection{Multichannel Audio}

The initial experiments aim to prove the effectiveness of multichannel audio in ASDL, as opposed to conventional visual face tracking followed by audio-visual ASD with single-channel audio.
Multichannel audio may not deliver good performance without adequate spatial processing or suitable network architecture.
Here, we study the network's temporal modeling and spatial input features, and relate these findings to systems without multichannel audio.
We first compare long and short input frames and test the use of recurrent units. 
An ablation study investigates spatial input features versus training the CRNN with log-mel spectrograms of the 16 microphone channels (\textsc{16mics}). 
To examine soundfield sampling, we compare with 1 and 2 channel log-mel spectrograms, i.e.,  (\textsc{Mono}) and (\textsc{Stereo}).
\textsc{Stereo} takes two microphones spaced at $\pm$88.3\,mm from the array center, consistent with the ORTF stereo microphone technique and approximating the spacing of human ears. 
\textsc{Mono} takes only the central microphone's signal.
Results are compared to those by single-channel audio-visual methods, \textsc{Asc} and \textsc{TalkNet}.

\subsubsection{Temporal Input and Modeling}

As in \cite{Cao:2019:polyphonic}, the audio input frames are extracted to have a fixed length of 2 seconds with a 1-second overlap for training.
We compare the student CRNN architecture trained with 2-second-long input frames with the architecture previously used in \cite{Berghi:2021:mmsp}. In \cite{Berghi:2021:mmsp}, the network processes a short input frame (167 ms) at a time with a similar architecture to the one proposed here without the recurrent units, i.e., convolutional layers followed by FC layers. We refer to it as `CNN-F', as per `frame-wise'. 
The CRNN architecture proposed in this study is also tested without the GRUs (`CNN') to compare the role of longer input frames on similar CNN-like architectures, as well as the benefits introduced by the recurrent units (`CRNN').

\subsubsection{Spatial Audio Input}

We extracted and tested two SOTA spatial input features from the multichannel audio signals: the GCC-PHAT \cite{Cao:2019:polyphonic} and SALSA-Lite \cite{Nguyen:2021:SALSALiteAF}. In each case, one of the $Ch_{in}$ channels is a spectrogram and the others are time-difference-based spatial representations. 
The GCC-PHAT between two audio channels, as in \cite{Cao:2019:polyphonic}, is calculated at every audio frame and represented as a matrix with time-lags on the frequency axis to allow concatenation with log-mel spectrograms from the array's channels. 
The maximum number of delayed samples corresponding to $\Delta\tau_{max}$ is computed as $d_{max}/c\cdot{f_s}$, where $f_s$ is the sampling frequency, $c$ the speed of sound, and $d_{max}$ the maximum distance between the two furthest microphones. The enable concatenation, the log-mel spectrogram's frequency resolution must be consistent with the number of time-lags. So, considering delay and advance between the signals, the number of mel-frequency bins must be greater or equal to $2\cdot{\Delta\tau_{max}}+1$ \cite{Cao:2019:polyphonic}. 
Since we only analyze the frontal horizontal domain and the speakers are always contained in the camera FoV, the formula can be modified as: $\Delta\tau_{max}=d_{rel}/c\cdot{f_s}$, with $d_{rel}$ representing the relative maximum distance between the microphones, given by $d_{max}\cdot{sin{\tfrac{\theta}{2}}}$, with $\theta$ being the camera horizontal FoV.
Cao \textit{et al.} compute the GCC-PHAT between all possible pairs of microphones in the array and append the log-mel spectrogram of each channel too. Each GCC-PHAT representation or log-mel spectrogram is a channel of the input tensor. However, we only compute the GCC-PHAT between a reference microphone and the remaining ones. We found that, for a large microphone array like the one of TragicTalkers, this approach not only produces significantly more compact inputs, but it also achieves better performance \cite{berghi:2023:WASPAA}. Additionally, since all microphones of the planar array face forwards, only one log-mel spectrogram from a single channel is computed.
Therefore, the first type of input feature includes a total of 16 channels: one log-mel spectrogram and 15 GCC-PHATs. For simplicity, we refer to them as GCC-PHAT features. 
SALSA-Lite features \cite{Nguyen:2021:SALSALiteAF} are a normalized version of the inter-channel phase difference (NIPD), computed at each time-frequency bin between a reference microphone and the remaining ones, and a log-linear spectrogram for each channel. As per GCC-PHAT, we only append a single log-linear spectrogram. Therefore, the second type of input features consists of 16 channels too.
In both GCC-PHAT and SALSA-Lite, we select the first microphone from the lower linear subarray, as highlighted in Fig.\,\ref{AVArig}, to be the reference microphone as tests conducted with the central microphone gave poorer performance.

In the TragicTalkers dataset, the video stream has resolution 2448$\times$2048p at 30\,fps, and audio is sampled at 48 kHz, 24 bits.
To align the output temporal resolution ($T_{out}=T_{in}/16$) with the labels frame rate, i.e., 60 activity-regression pair predictions for the 2-second audio input,
we apply an STFT with Hann window of size 512 samples at hop steps of 100 samples. Thus, the 2-second (96k-sample) audio chunk is discretized into 960 temporal bins ($T_{in}$).
To compute the log-mel spectrogram used for the GCC-PHAT features, the frequency resolution of the spectrograms is down-sampled over 64 mel-frequency bins and the logarithm operation is then applied. 
Also, the distance between the two furthest microphones in the array, $d_{max}$, is 450 mm and the camera's horizontal FoV, $\theta$, is roughly 55°. Therefore, 64 time-lags are just enough to concatenate the GCC-PHAT `spectrograms' with the single-channel log-mel spectrogram.
For the log-linear spectrogram and the NIPD features used for SALSA-Lite, due to the limited presence of speech information at higher frequencies and to avoid spatial aliasing, an upper cutoff frequency of 6kHz is used. 
This extracts the first 64 frequency bins to be consistent with the input shape of GCC-PHAT. The logarithm operation is then applied.

All input features are normalized for zero mean and unit standard deviation vectors, frequency-wisely and channel-wisely.
That is, for each input channel, two $F_{in}$-long vectors are computed from the training set, representing respectively the mean and the standard deviation of each frequency bin (or time-lag bin in the GCC-PHAT). 
In the original SALSA paper \cite{Nguyen:2021:SALSA}, only the spectrogram channels were normalized in such a way, but with this microphone array better results are achieved by normalizing all the input channels.

\subsection{Supervision}

We conducted a series of experiments to investigate the supervision of the proposed student-teacher learning pipeline. In particular, we compared the CRNN trained following our self-supervised method and the CRNN trained under the full supervision of the GT labels. 
As anticipated, we tested both \textsc{Asc} and \textsc{TalkNet} to represent respectively weak and strong teacher networks.
Since for the regression subtask the student network is penalized only when the positional pseudo-labels are available, the main problem with a weak teacher network primarily concerns false positive (FP) detections, i.e., when the silent actor is incorrectly classified as active. These FP predictions introduce noise in the set of pseudo-labels used for training.
In order to quantify the impact these noisy labels might cause, we manually screened the predictions of \textsc{Asc} to remove all FP predictions. The screened version, 
\textsc{Asc(s)}, is compared to the original \textsc{Asc} teacher network.
Furthermore, since the supervision is based on two separate sets of automatically generated labels (the positional pseudo-labels and the VA labels), we also tested hybrid versions of manual-automatic supervision. 
Leveraging the GT labels of the dataset, we combined the pseudo-labels generated by the teacher networks with the GT VA labels. Similarly, we tested the GT positional labels with the VA labels generated with the VAD. As a VAD we adopted the open-source WebRTC \cite{google:webRCT} as it is reliable and widely employed.
In total, this yields 8 combinations, as summarized in Tab.\,\ref{tab:supervMatrix}.

To better understand the different types of supervision compared in this study, we present a rough estimate of the number of hours required to manually annotate a dataset such as TragicTalkers with 30 scenes (640s total) captured by two microphone arrays and 22 cameras in total.
\begin{itemize}
    \item \textbf{Voice Activity:} It requires on average about 10-15 minutes to annotate the onset and offset times of each speech segment in a sequence. Since this can be completed once and applied for both arrays, it therefore takes 5-7.5 hours to label the entire dataset. $\sim$\,40:1 real-time (RT) on audio.
    \item \textbf{False positive screening:} Removing FP predictions from the output of an ASD takes roughly 5 minutes per camera view. Yet, single-speaker sequences (monologues) often only require minor corrections. So,  TragicTalkers' 15 two-speaker sequences take more than 27 hours of labeling work. $\sim$\,7:1 RT on video.
    \item \textbf{Bounding boxes:} Even with the help of a face detector, manually drawing the bounding boxes around the face of the actors when the detector fails, requires 10 to 15 minutes per camera view, yielding up to 165 hours of labeling work. $\sim$\,40:1 RT on video.
\end{itemize}
In addition to that, rest breaks to reduce fatigue of the labeler have to be taken into account, especially during the bounding box labeling task. 
Although this only represents a rough estimate of the labeling times and additional factors should be taken into account, e.g., the experience of the labeler, it emphasizes how costly manual annotation can be.
Considering the types of supervision that we study in Tab.\,\ref{tab:supervMatrix}, a `Spectrum of Supervision Conditions' is proposed in Fig.\,\ref{Fig:supLine}, with the fully supervised case on the left end and the fully self-supervised on the right end. In between are reported the hybrid combinations, sorted from the less labor-intensive to the most on TragicTalkers.


\begin{table}[tb]
\caption{Supervisory Conditions. Columns report the source of location pseudo-labels, rows the voice activity labels (VA). Naming convention as [Location]-[VA] labels.}
\begin{center}
\begin{tabular}{c|c|c|c|c|c|}
\multicolumn{2}{c}{} & \multicolumn{4}{c}{\textbf{Face Detection and Localization}} \\
\cline{3-6}

\multicolumn{2}{c|}{\mbox{ }\hspace{-3ex}\textbf{VA}\hspace{-3ex}\mbox{ }} & \textbf{{Supervised}} & \textbf{{Screened}} 
& \multicolumn{2}{c|}{\textbf{{Weak}\,/\,{Strong} Teacher}} \\
\cline{2-6}

\multicolumn{2}{|c|}{\textbf{\textsc{Gt}}} 
& \mbox{\scriptsize (1)}\,\textsc{Gt-Gt} 
& \mbox{\scriptsize (3)}\,\textsc{Asc(s)-Gt} 
& \multicolumn{2}{c|}{\mbox{\scriptsize (5)}\,\textsc{Asc-Gt}\,/\,\textsc{TalkNet-Gt}} \\
\cline{2-6}

\multicolumn{2}{|c|}{\textbf{\textsc{Vad}}} 
& \mbox{\scriptsize (2)}\,\textsc{Gt-Vad} 
& \mbox{\scriptsize (4)}\,\textsc{Asc(s)-Vad} 
& \multicolumn{2}{c|}{\mbox{\scriptsize (6)}\,\textsc{Asc-Vad}\,/\,\textsc{TalkNet-Vad}} \\
\cline{2-6}

\end{tabular}\vspace{-3ex}
\label{tab:supervMatrix}
\end{center}
\end{table}

\begin{figure}[tb]
\centerline{\small\textsc{Gt-Gt}\hfill\textsc{Asc(s)-Gt}\hfill\textsc{Teacher-Gt}\hspace{15mm}\mbox{ }}
\centerline{\small\circled{1}\rule{11mm}{0.1pt}\circled{2}\rule{11mm}{0.1pt}\circled{3}\rule{11mm}{0.1pt}\circled{4}\rule{11mm}{0.1pt}\circled{5}\rule{11mm}{0.1pt}\circled{6}}
\centerline{\small\mbox{ }\hspace{15mm}\textsc{Gt-Vad}\hfill\textsc{Asc(s)-Vad}\hfill\textsc{Teacher-Vad}}
\caption{Spectrum of Supervision Conditions. 
From fully supervised (left) to fully self-supervised (right), where
\mbox{\small\textsc{Teacher}}=\{\mbox{\small\textsc{Asc}},\mbox{\small\textsc{TalkNet}}\}}
\label{Fig:supLine}
\end{figure}

\subsection{Evaluation Metrics}

A 5-fold sequence-wise cross-validation approach is implemented to train the audio network: each validation fold sets aside 10 unseen sequences from the 50 sequences of the development set. This cross-validation approach is used to find suitable hyperparameters for the network. Once found, the model is retrained using the entire 50-sequence training set with these values.
The network is trained for 50 epochs using batches of 32 audio feature inputs and Adam optimizer. The learning rate is fixed for the first 30 epochs, then reduced by 10\% each epoch. 
The initial learning rate determined in the cross-validation is typically \num{e-4} but varies 
according to the supervision condition.

We evaluate our method on the 10-sequence test set of TragicTalkers, in which each camera view is labeled for 2D mouth position, making 110 test sequences in combination (44min). The ground truth mouth coordinates correspond to the mouth keypoint extracted with OpenPose \cite{Cao:2019:openpose}. They were manually checked to ensure that the keypoint lay within the mouth area, which is,\,$\sim$25 pixels-wide, depending on distance ($\pm$\,0.3°).  
A frame prediction is considered to be positive, i.e. the network predicts the presence of speech, when the confidence is above a threshold and a positive detection is considered to be true when the localization error is within a predefined tolerance. 
The precision and recall rates are computed by varying the confidence threshold from 0\% to 100\%. Since, at inference, the confidence of the network tends to be either very high or very low, the precision-recall curves are built by sampling the thresholds from a Sigmoid-spaced distribution. This provides more data points for high and low confidence values. 
The popular object detection metric average precision (AP) was then computed. The AP is determined following the approach indicated by the Pascal VOC Challenge \cite{Everingham:2015:pacalVOC}: (1) compute a monotonically decreasing version of the precision-recall curve by setting the precision for the recall \textit{r} equal to the maximum precision obtained for any recall \textit{r'$\geq$r}, and (2) compute the AP as the numerical integration of the curve, i.e. the area under the curve (AUC). 

According to human auditory spatial perception \cite{Strybel:2000:MinimumAA}, the minimum audible angle (MAA) is 2°. Therefore, a tolerance for spatial misalignment between the prediction and the GT speaker's position of $\pm$2° along the azimuth is set, which corresponds to $\pm$89 pixels projected onto the image plane,
based on camera-calibration data.
In many AV applications, the human brain can accommodate wider misalignments \cite{Stenzel:2018:PTC}. 
Thus, the AP is computed at $\pm$5° tolerance ($\pm$222 pixels) too. 
The F1 score is computed from the precision-recall pairs at the optimal compromise between precision and recall rates,
yielding a summary metric of detection and localization for comparing methods. 
The average distance (aD) and detection error (Det Err \%) are employed to assess performance in the localization and detection subtasks, respectively. 
The aD only takes into account the true positive (TP) detections and it represents the average distance in pixels between the active detections and GT locations. We also report the respective angular localization errors by converting pixels to degrees to facilitate the comparison with other sound localization works.
Det Err quantifies incorrect active-silent classifications, as a percentage, with confidence threshold set at 0.5, regardless of regression accuracy and spatial tolerance.
The metrics are computed on the overall test set but also separately for each of the 10 test sequences across both AVA rigs. Hence, the 10 values enable  statistical significance tests.







\section{Experimental Results and Discussion} \label{sec:results}

First, we present the results of the temporal modeling study, then, the ablation study switching on and off the spatial feature extraction and reducing the number of microphones. The results of the multichannel audio methods are compared to the performance of the existing AV ASD. Additionally, to test the robustness of the audio network, we re-trained the model corrupting the dataset with additive pink noise at SNRs from 0 dB to 40 dB.
Finally, we discuss the results achieved under various supervision conditions.
We conclude the section discussing the limitations of the proposed approach and proposing directions for future research.

\subsection{Multichannel ASDL}

\subsubsection{Temporal Input and Modeling}

For each of the three networks tested, the training was performed with the \textsc{Gt-Gt} labels and the GCC-PHAT input features. Results are reported in Tab.~\ref{tabArchitecture}. 
CNN produces a remarkable improvement in detection error compared to CNN-F. This can be partially attributed to the length of the input frames and partially to the employment of VA labels instead of the silent sequence to generate negative samples. 
This improvement is reflected in an increment of 8 percentage points in AP@2°, and about 0.05 in F1@2°. The introduction of the GRUs with longer audio frames does not improve the aD performance (39p (0.9°) in both cases). However, there is a benefit in the detection error with over 1.5 percentage points improvement between CNN and CRNN.
Therefore, with the same aD but a lower detection error, CRNN also enables greater AP and F1 score.
These results suggest that longer temporal audio horizons provide the network with useful information to track the target speaker over time achieving higher spatial accuracy. However, it does not seem to improve the detection performance. 
When it comes to partnering more extended frames with recurrent units, the benefit affects the detection subtask too. This suggests that learning the recurrent temporal dependency between the time bins facilitates activity detection, providing appropriate conditions to better tackle the overall ASDL task.

%
\begin{table}[tb]
\caption{Temporal Modeling. Detection error (DetErr) and average distance (aD); average precision (AP) and F1 score at 2° ($\pm$89p) tolerance.
Results 
with GCC-PHAT input features and fully supervised training (Gt-Gt).}
\begin{center}
\begin{tabular}{c|c|c|c|c|c}
\hline
\textbf{Method}&\textbf{Input\,len}&\textbf{DetErr}&\textbf{aD}&\textbf{AP@2°}&\textbf{F1@2°} \\  \hline
CNN-F\,\cite{Berghi:2021:mmsp} & 167\,ms & 13\% & 42p,\,0.95° & 78\% & 0.850 \\
CNN & 2000\,ms & 4.7\% & \textbf{39p,\,0.88°} & 86\% & 0.902 \\
CRNN & 2000\,ms & \textbf{3.2\%} & \textbf{39p,\,0.88°} & \textbf{87\%} & \textbf{0.909} \\  \hline 
\end{tabular}\vspace{-3ex}
\label{tabArchitecture}
\end{center}
\end{table}

\begin{table}[tb]
\caption{Modality Potential. Input `A': audio-only; `AV': audio-visual; `-S': single channel; `-M': multichannel. 
For AV, predictions are the detected faces' bounding box center.
For A, methods trained with ground truth labels (Gt-Gt).}
\begin{center}
\begin{tabular}{c|c|c|c|c|c}
\hline
\textbf{Method}&\textbf{Input}&\textbf{DetErr}&\textbf{aD}&\textbf{AP@2°}&\textbf{F1@2°} \\ \hline
\textsc{Mono} & A-S & 2.7\% & 210p,\,4.7° & 10\% & 0.300  \\
\textsc{Stereo} & A-M & 2.6\% & 100p,\,2.3° & 31\% & 0.535  \\
\textsc{16mics} & A-M & \textbf{2.5\%} & 67p,\,1.5° & 56\% & 0.710 \\  \hline
\textsc{Gcc-Phat} & A-M & 3.2\% & 39p,\,0.88° & \textbf{87}\% & \textbf{0.909}  \\
\textsc{Salsa-Lite} & A-M & 3.6\% & 40p,\,0.90° & 85\% & 0.895 \\  \hline
\textsc{Asc}\,\cite{Alcazar_2020_CVPR} & AV-S & 43\% & 50p,\,1.1° & 59\% & 0.676  \\
\textsc{Asc(s)} & AV-S & 24\% & \textbf{23p,\,0.52°} & 62\% & 0.704 \\
\textsc{TalkNet}\,\cite{tao2021TalkNet} & AV-S & 14\% & 35p,\,0.79° & 82\% & 0.849 \\  \hline

\end{tabular}
\label{tabASDL}
\end{center}
\end{table}

\subsubsection{Spatial Audio Input}
  
The \textsc{Gcc-Phat} and \textsc{Salsa-Lite} methods extract spatial features from multichannel audio input. 
The two {middle} rows of Tab.\,\ref{tabASDL} present their results, alongside those for the log-mel spectrograms extracted directly from the array's microphones, as \textsc{Mono}, \textsc{Stereo} and \textsc{16mics}.
All these audio-based methods are trained in a fully supervised way with the \textsc{Gt-Gt} labels combination.
In the \textsc{Mono} method, the detection subtask is easily accomplished with a detection error of only 2.7\%. However, the absence of spatial cues does not allow the correct regression of the position of the speaker. In fact, to minimize the error, the model tries to locate the speaker in the central area of the frame. By adding multiple channels, the spatial accuracy improves significantly. \textsc{Stereo} achieves less than half the aD achieved by \textsc{Mono} and it remarkably increases its F1@2°. With all 16 microphones, this gain is further amplified: the aD is a third of the one achieved in \textsc{Mono} and the F1@2° is almost 2.5 times bigger.
Despite that the employment of multiple microphones improves spatial accuracy, \textsc{Stereo} and \textsc{16mics} methods are still poor in terms of AP and F1 score when compared to \textsc{Gcc-Phat} or \textsc{Salsa-Lite}. This suggests that the direct use of the array channels without spatial input extraction is not particularly effective since important directional cues encoded in the time difference of the signals are neglected. In contrast, the employment of input feature extraction enables the achievement of high performances. \textsc{Salsa-Lite} produces an increment in F1@2° of 0.185 points compared to \textsc{16mics}, while \textsc{Gcc-Phat} 0.199 points. 
This indicates that spatial feature extraction provides great benefits in solving the ASDL problem. Most of the improvement is a consequence of the remarkably higher spatial accuracy: in terms of detection errors, the methods without spatial feature extraction achieve slightly better results.

\begin{figure}[tb]
\centerline{\includegraphics[width=\columnwidth]{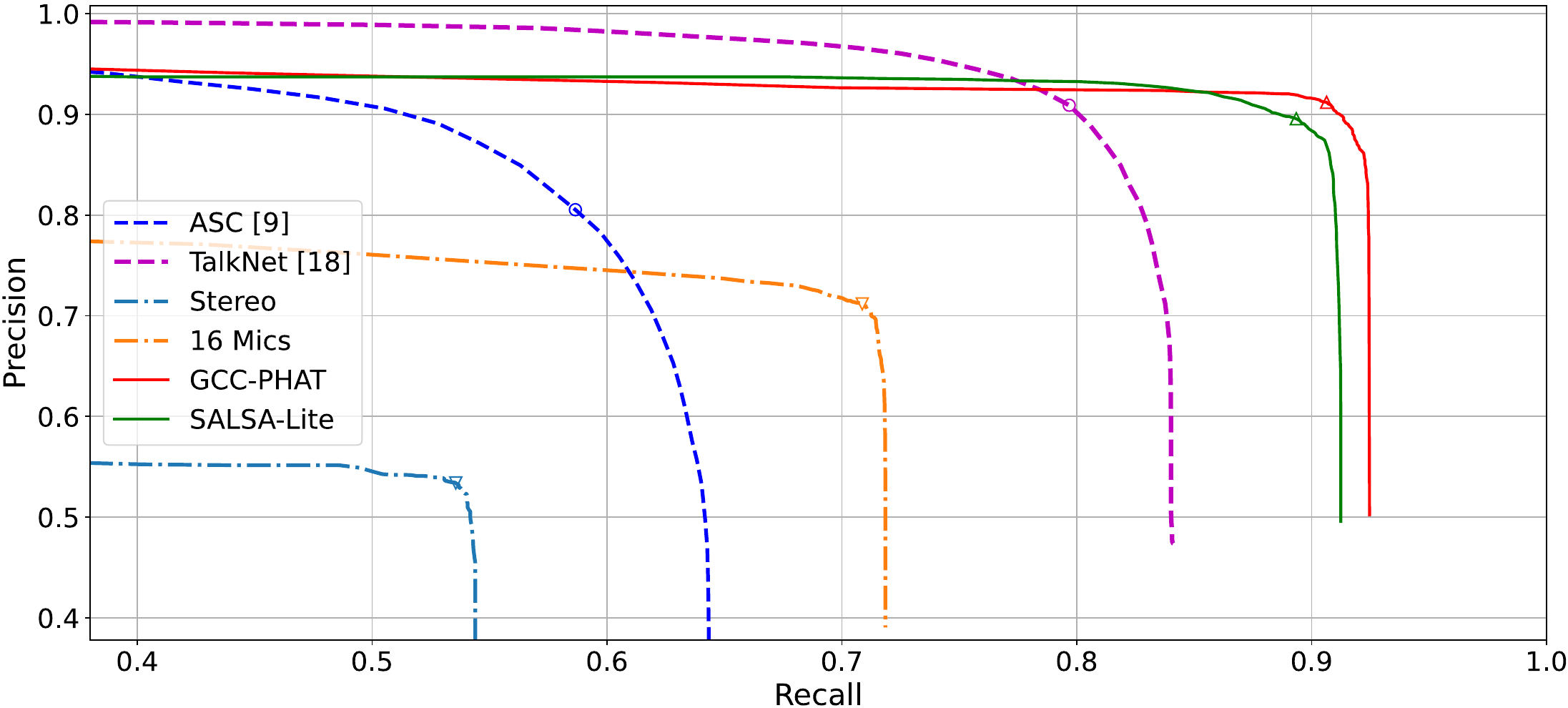}}
\caption{Audio-visual ASD and audio-only ASDL methods comparison of precision versus recall at 2° tolerance (89 pixels). The combination of precision and recall rates that achieves the highest F1 score is marked on each curve.}
\label{Fig:precision_recall}
\end{figure}

\subsubsection{Comparison with Existing Active Speaker Detectors}

Conventional active speaker detectors, such as \textsc{Asc} and \textsc{TalkNet}, employ the audio modality only to classify the pre-extracted faces. 
The localization subtask is performed by the visual face detector, yielding high spatial accuracy.
However, the average distance is penalized by the FP detections of the classifier. In other words, the aD values reported in Tab.\,\ref{tabASDL} for \textsc{Asc} and \textsc{TalkNet} are influenced by the silent actor being misclassified as active. In practice, this problem only affects \textsc{Asc}, as in \textsc{TalkNet} the silent speaker is almost never misclassified.  
To understand to what extent the aD of \textsc{Asc} is penalized by these episodes, one can observe the aD achieved by \textsc{Asc(s)}. The aD is halved to 23p (0.5°) and the detection error is remarkably decreased too, suggesting that roughly half of the detection error affecting \textsc{Asc} is caused by FP predictions, while the other half by misdetections, i.e., when the face detector fails in detecting the face of the speaker due to visual occlusions or self-occlusions. 
The even lower detection error achieved by \textsc{TalkNet} indicates that its face detector (S3FD \cite{Zhang2017S3FDS}) detected more active faces than SeetaFaceEngine2 \cite{wu2016Seeta}, used in the \textsc{Asc} method.   
In the AV methods, the horizontal coordinate of the center of the bounding box is used as prediction, while the ground truth used for evaluation refers to the actual mouth position of the speaker. Therefore, the already low aD achieved might be even slightly overestimated due to the offset between the two representations. For example, when the speaker is captured in profile and his/her mouth is closer to the edge of the bounding box.
The fairly high detection error achieved by the AV active speaker detectors caused by the visual misdetections is reflected in a gap in terms of recall rate, as it can be appreciated from Fig.\,\ref{Fig:precision_recall}, especially for \textsc{Asc}. At their best precision-recall pair, with 2° tolerance, \textsc{TalkNet} present a recall rate of 79.7\%, while \textsc{Asc} of 58.5\% only. 
As a consequence, their overall F1 scores are affected too.
As previously mentioned, the reason why \textsc{Asc} does not perform so well on TragicTalkers relates to the training on the AVA-ActiveSpeaker dataset where the faces are extracted with a different algorithm. 

In contrast, detection represents a relatively trivial task for the multichannel audio methods that can sense and detect speech activity even when the speaker is visually occluded. In fact, the double-digit detection errors of the AV methods are considerably reduced. Thus, for the multichannel audio methods, the F1 score mainly depends on their localization accuracy. 
Fig.\,\ref{Fig:precision_recall} shows how the gap in recall rate generated by \textsc{TalkNet} is halved with the \textsc{Gcc-Phat} and \textsc{Salsa-Lite} methods (90.6\% for \textsc{Gcc-Phat} and 89.4\% for \textsc{Salsa-Lite}). This produces an AP and F1 score higher than the other AV systems that employ single-channel audio. The F1 score achieved by \textsc{Gcc-Phat} is significantly greater than the one achieved by \textsc{TalkNet} ($p=0.02$).
The active frames that \textsc{Gcc-Phat} and \textsc{Salsa-Lite} do not detect are mainly caused by wide predictions, where speech activity was correctly identified but the estimated locations were outside the tolerance. Their aD is almost twice as big as the one visually achieved by \textsc{Asc(s)}. In fact, if the tolerance threshold is increased to 5°, \textsc{Salsa-Lite} achieves an AP@5° of 99.4\% and an F1@5° of 0.971, while \textsc{Gcc-Phat} similarly yields an AP@5° of 98.9\% and an F1@5° of 0.974. So, with the broader 5-degree tolerance, the residual error is under 3\% and predominantly attributable to the detection errors.

\subsubsection{Robustness to Additive Noise}

Since the controlled studio conditions of the dataset are more benign than many other application contexts, we tested the audio network's robustness, corrupting the TragicTalker dataset with additive pink noise at SNRs from 0\,dB to 40\,dB and re-training the model. Results achieved with \textsc{Gt-Gt} labels are presented in Tab.\,\ref{tabPinKNoise}. Note how performance improves in clean conditions and high-SNR cases compared to our results when training only on clean data. This is because adding noise is a form of data augmentation that artificially increases the size of the dataset, suggesting that further gains may be obtained with larger datasets and other forms of augmentation. The network proved to be robust with an F1@2°\,$>$\,0.9 for SNR\,$\geq$\,20\,dB.
Under heavy noise conditions, both detection and localization accuracy are slightly penalized. Yet, the detection error remains well below those from the audio-visual teacher networks; the F1 score exceeds 0.93 with the 5-degree threshold at the lowest 0\,dB SNR tested.

\begin{table}[tb]
\caption{Robustness to additive pink noise.
Results achieved with GCC-PHAT features and fully supervised training (Gt-Gt).}
\begin{center}
\begin{tabular}{c|c|c|c|c|c}
\hline
\textbf{SNR}&\textbf{DetErr}&\textbf{aD}&\textbf{AP@2°}&\textbf{F1@2°}&\textbf{F1@5°} \\  \hline
\textbf{Clean} & \textbf{2.6\%} & \textbf{37p,\,0.83°} & \textbf{89\%} & \textbf{0.930} & \textbf{0.980} \\
\textbf{40\,dB} & 2.9\% & \textbf{37p,\,0.83°} & \textbf{89\%} & 0.929 & 0.977 \\
\textbf{30\,dB} & 2.8\% & \textbf{37p,\,0.83°} & 88\% & 0.924 & 0.978 \\ 
\textbf{20\,dB} & 3.3\% & 39p,\,0.88° & 86\% & 0.909 & 0.974\\
\textbf{10\,dB} & 4.0\% & 40p,\,0.90° & 83\% & 0.890 & 0.967\\
\textbf{0\,dB} & 6.9\% & 51p,\,1.15° & 70\% & 0.782 & 0.932\\ \hline 
\end{tabular}\vspace{-3ex}
\label{tabPinKNoise}
\end{center}
\end{table}

\begin{table}[tb]
\caption{Supervision results 
with F1 reported at 2° and 5° tolerances.}
\begin{center}
\begin{tabular}{c|c|c|c|c|c}
\hline
\textbf{Supervision}&\textbf{DetErr}&\textbf{aD}&\textbf{AP@2}&\textbf{F1@2}&\textbf{F1@5} \\  \hline
\textsc{Gt-Gt} & 3.2\% & \textbf{39p,\,0.88°} & \textbf{87\%} & \textbf{0.909} & \textbf{0.975}  \\
\textsc{Gt-Vad} & 7.1\% & \textbf{39p,\,0.88°} & 85\% & 0.878 & 0.938  \\  \hline
\textsc{Asc(s)-Gt} & 3.5\% & 42p,\,0.95° & 83\% & 0.881 & 0.970  \\
\textsc{Asc(s)-Vad} & 6.6\% & 43p,\,0.97° & 81\% & 0.844 & 0.936  \\  \hline
\textsc{Asc-Gt} & \textbf{3.0\%} & 110p,\,2.5° & 39\% & 0.606 & 0.790  \\
\textsc{Asc-Vad} & 7.0\% & 100p,\,2.3° & 38\% & 0.594 & 0.787  \\  \hline
\textsc{TalkNet-Gt} & 3.6\% & 43p,\,0.97° & 85\% & 0.890 & 0.970   \\
\textsc{TalkNet-Vad} & 7.5\% & 42p,\,0.95° & 81\% & 0.854 & 0.935  \\  \hline
\end{tabular}
\label{tabSupervision}
\end{center}
\end{table}

\subsection{Spectrum of Supervision Conditions}

The results achieved by the student network trained with the different types of supervision are reported in Tab.\,\ref{tabSupervision}.
Since the network trained with the GCC-PHAT input features in the previous study achieved a higher F1 score, the results reported in the table are also obtained using the GCC-PHAT. Nevertheless, we did not observe statistical significance of the GCC-PHAT being better than SALSA-Lite ($p=0.08$). 

\subsubsection{Effects of the VAD}

The employment of the VAD penalizes detection accuracy. The student network trained using the GT regression labels and the VAD activity labels (\textsc{Gt-Vad}) generates a detection error of 7.1\%, compared to the 3.2\% achieved by \textsc{Gt-Gt}. This is also reflected in a decrement of about 0.031 points in F1@2° ($p<0.01$).
The same trend can be appreciated with the other teacher networks in Tab.\,\ref{tabSupervision}: the employment of the VAD labels generates increments in detection error of 3-4 percentage points.
This behavior was predictable since the GT VA labels in the development and test sets are manually generated with the same labeling policy and therefore the predictions lead to smaller detection errors. 
In contrast, the VAD is less precise in detecting the onsets and offsets of the speech segments.
Nevertheless, supervising the student network with the VAD labels does not seem to substantially affect localization accuracy. 

\subsubsection{Comparison of Teacher Networks}

\begin{figure}[tb]
\centerline{\includegraphics[width=\columnwidth]{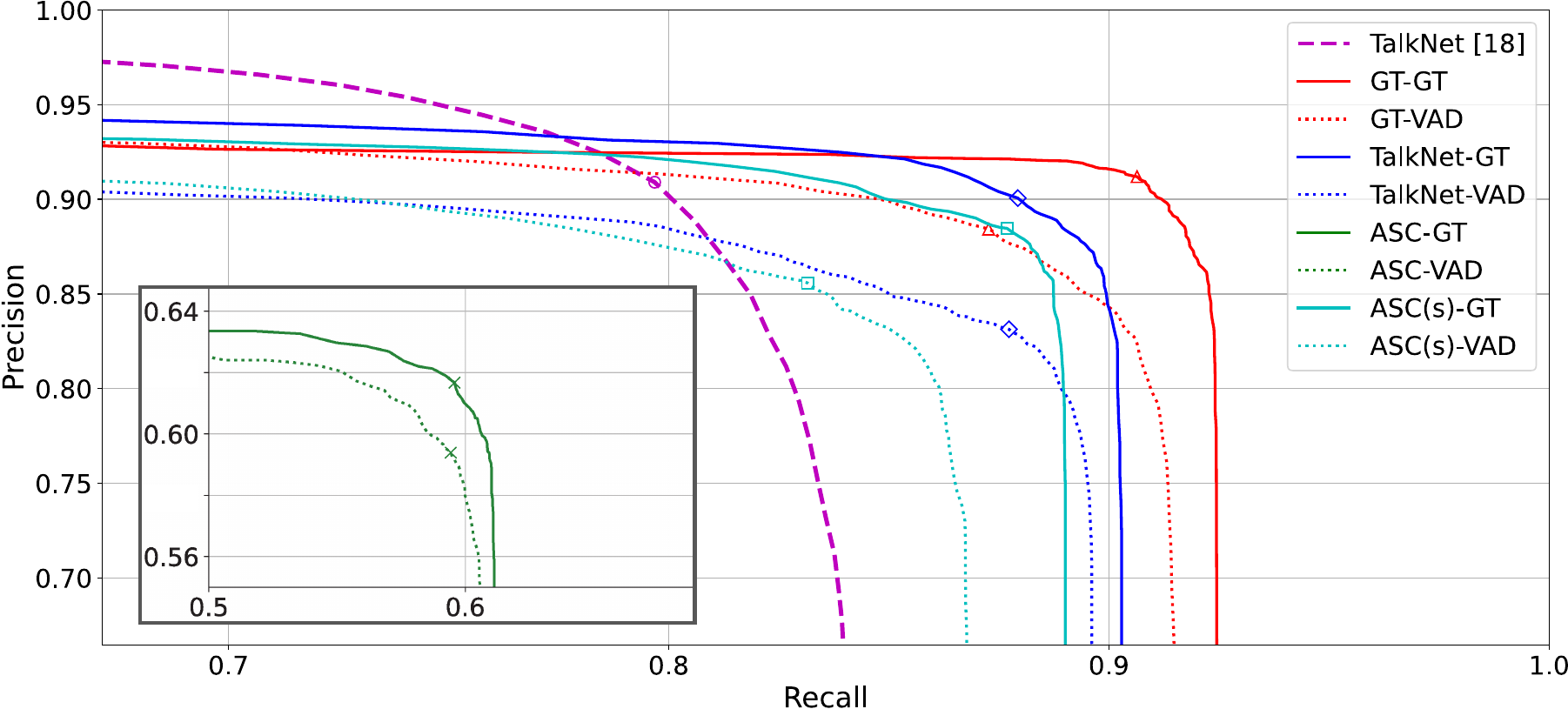}}
\caption{Comparison of precision versus recall at 89-pixel tolerance (2°) for the different supervision methods. The curve achieved with the \textsc{TalkNet} method is included for reference. The combination of precision and recall rates that achieves the highest F1 score is marked on each curve.}
\label{Fig:precision_recall_sup}
\end{figure}

From Tab.\,\ref{tabSupervision}, it is clear that \textsc{Asc} provides poor supervision due to the noise in its pseudo-labels caused by the FP detections, yielding students' aD results about 2.5 times larger than the student networks trained with the strong teacher. This deficiency is particularly evident also in Fig.\,\ref{Fig:precision_recall_sup}, which reports the precision-recall curves achieved with the different supervisory conditions as well as the one from \textsc{TalkNet} for comparison. 
In fact, when it comes to supervising the student network with \textsc{Asc(s)}, the aD remarkably improves and is almost comparable to the one achieved in the fully supervised case (\textsc{Gt-Gt}). Actually, with the SALSA-Lite input features, \textsc{Asc-Gt} marginally outperforms \textsc{Gt-Gt} with an F1@2° of 0.901 against the 0.895 achieved with full supervision.
This is because when the pseudo-labels do not present FP detections thanks to manual screening, they are analogous to the regression labels used as ground truth as they both use ``clean'' bounding box coordinates. The main difference is the quantity: in training, \textsc{Asc(s)} only uses the visible faces while the GT labels also include the frames where the face is not visible.
As mentioned previously, \textsc{TalkNet} does not produce many FP detections. With the GCC-PHAT features, the student network trained under the supervision of \textsc{TalkNet-Gt} is the one that achieved the highest AP and F1 score at 2° and 5°, being outperformed only by the fully supervised case.
\textsc{Asc-Vad} and \textsc{TalkNet-Vad} are the only two combinations enabling fully self-supervised training as they do not involve manual labeling or screening.
In particular, the results achieved with \textsc{TalkNet-Vad} prove that a competitive ASDL system can be trained in a fully self-supervised fashion. Furthermore, the student network trained with \textsc{TalkNet-Vad} outperformed its own teacher with an F1 score respectively of 0.854 and 0.935 at 2° and 5°, compared to the 0.849 and 0.855 achieved by the \textsc{TalkNet} method. With SALSA-Lite, the F1 score is even higher achieving 0.860 at 2° and 0.941 at 5°.
The spatial accuracy achieved by the \textsc{TalkNet} method is yet more efficient thanks to the visual modality. This is also reflected in greater levels of precision rate in the earlier part of its precision-recall curve corresponding to high confidence levels. 

These results suggest that achieving good performance with the proposed learning pipeline is possible, as long as a good teacher network is adopted.
In addition to a good teacher network to provide a reliable set of pseudo-labels, even better results with minimal manual supervision can be achieved by labeling the audio stream for VA. In fact, the student network trained with \textsc{TalkNet-Gt} is the one that achieved the highest ASDL performance with GCC-PHAT features, only the fully supervised methods obtained better results.
When a good teacher network is not available, like in the case of \textsc{Asc}, a useful strategy to maintain competitive results would be to manually screen the FP detections of the teacher. However, in a multi-view dataset like TragicTalkers, this process requires a more labor-intense labeling process and does not achieve results as good as the ones achieved with a better teacher and manual VA annotation.  
Nevertheless, all the hybrid solutions tested represent a viable alternative to traditional supervised learning as they allow for saving a great amount of labeling time and resources with little sacrifice in terms of performance.

\subsection{Limitations and Future Directions}

Experimenting on TragicTalkers has enabled rigorous assessment of the proposed learning approach under controlled studio conditions and with the simple single-speaker scenario, which covers the vast majority of professionally produced media content. 
Nonetheless, different scenarios may present external environmental noises or multiple simultaneous talkers. We partially addressed this problem by testing various levels of additive noise. However, this still does not cover realistic in-the-wild situations. Further experiments should be conducted in this direction to comprehensively study the model's robustness for these settings. 
Furthermore, although a similar supervision method can in theory be applied for multiple simultaneous speakers, 
additional design decisions, implementation and testing are needed.
To provide full supervision, the teacher network will need to generate pseudo-labels for each of the active speakers in a mixture of voices, which may be facilitated practically with close microphones or synthetically by making artificial mixtures.

The current supervision approach is spatially constrained by the camera FoV. 
From an audio perspective, this is limiting as sound can be sensed from all directions. 
To provide visual supervision that covers a larger range, a wide-angle or a 360° camera can be employed instead. Future work will investigate audio-visual systems able to detect and localize speakers both within and outside the FoV.

\section{Conclusion} \label{sec:concl}

In this paper, a student-teacher learning approach is proposed to tackle the ASDL task via a multichannel CRNN audio network.
The proposed learning pipeline is based on positional pseudo-labels generated with existing audio-visual active speaker detectors and VA labels generated with a VAD.
Experiments conducted to compare CRNN and CNN architectures suggest that recurrent units and longer input frames provide the network with important contextual information to better tackle the ASDL problem.
An ablation study to prove the importance of spatial input feature extraction found that it slightly penalizes the detection accuracy but improves up to 40\% the spatial accuracy. Therefore, it represents a beneficial practice for ASDL.

Extensive studies on the TragicTalkers dataset found that the CRNN with multichannel audio outperforms by a wide margin the AV methods with single-channel audio, especially in recall rate. Furthermore, tests with additive pink noise demonstrated the robustness of the audio network in adverse conditions.
It achieves better F1 scores than the AV ASD methods because more reliable voice activity detection gives higher recall rates than visual methods, which suffer from facial occlusions.  
The same detection advantage explains why the proposed student-teacher pipeline is successful at the ASDL task: during training, the student network is tuned only to the teacher's active detections for the localization subtask (the visible speakers); at inference, its focus on audio information allows it to generalize for visually-occluded speakers.
The fully self-supervised student network achieved 0.854 F1@2°, outperforming its own strong teacher network that achieved 0.849.

Studies with various supervisory conditions demonstrated the effectiveness of the proposed fully self-supervised learning pipeline, as long as a strong teacher network is employed. The performance can be further improved with minimal manual supervision. 
For example, ground truth VA labels increased the F1@2° to 0.890.
An advantage of the GT localization labels, as opposed to those of a strong teacher network, is the quantity: GT guarantees tracking data even when the speaker is visually occluded. The limitations of the current approach and directions for future research have been discussed.
For example, the proposed learning pipeline can be applied to larger datasets in order to assess whether the gap between fully supervised and other forms of supervision proposed can be further reduced.
We will explore an audio-visual system for ASDL with multichannel audio to improve spatial accuracy while preserving the high detection rates provided by the multichannel audio, taking into account problems like simultaneous multi-talkers and off-screen speakers.




\section*{Acknowledgments}

The authors would like to thank Mark Plumbley for his helpful comments.
This research was supported by InnovateUK (105168) `Polymersive: Immersive video production tools for studio and live events', UKRI EPSRC and BBC Prosperity Partnership AI4ME: Future Personalised Object-Based Media Experiences Delivered at Scale Anywhere EP/V038087 led by Adrian Hilton, and a Doctoral College PhD studentship at the University of Surrey. For the purpose of open access, the author has applied a Creative Commons Attribution (CC BY) licence to any Author Accepted Manuscript version arising. Data supporting this study are available from \url{https://cvssp.org/data/TragicTalkers} (DOI: 10.15126/surreydata.900446).



\bibliographystyle{IEEEtran}
\bibliography{TASLP2023}


 

\begin{IEEEbiography}[{\includegraphics[width=1in,height=1.25in,clip,keepaspectratio]{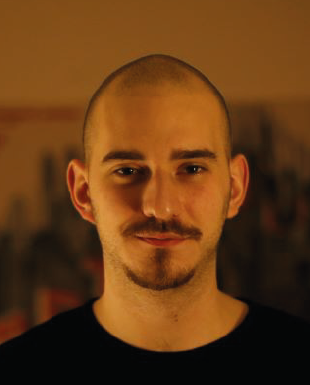}}]{Davide Berghi}
is pursuing a PhD at the Centre for Vision, Speech \& Signal Processing (CVSSP, University of Surrey, UK). He received his MSc in Computer Vision, Robotics, and Machine Learning from the same university, and BSc in Electronics and Telecommunication Engineering from the Dep. of Information Engineering and Computer Science (DISI. University of Trento, IT) in 2019 and 2017, respectively. 
His research focuses on audio-visual active speaker detection and localization. He won the best student paper award at IEEE MMSP2021. 
\end{IEEEbiography}

\begin{IEEEbiography}
[{\includegraphics[width=1in,height=1.25in,clip,keepaspectratio]{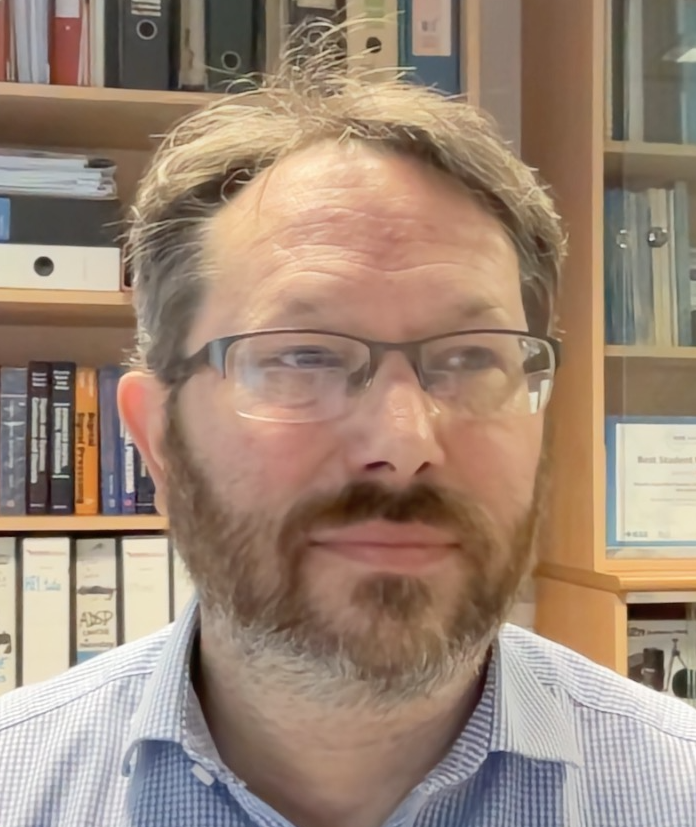}}]{Philip Jackson}
received his B.A. in Engineering from Cambridge University and Ph.D. from the University of Southampton, UK. He joined the Centre for Vision, Speech \& Signal Processing (CVSSP, University of Surrey, UK) in 2002, where he is Professor in Machine Audition, researching topics from immersive audio and audio-visual perception to sound zones and object-based media [scholar: \url{https://tinyurl.com/soundspatially}, h-index 30].

\end{IEEEbiography}

\vfill

\end{document}